\documentclass[10pt,aps,prd,notitlepage,amssymb,amsmath,floatfix,nofootinbib,superscriptaddress]{revtex4-1}
\usepackage{epsfig}
\usepackage{bm}
\usepackage{amssymb}
\usepackage{amsmath}
\usepackage{color}
\usepackage{xcolor}
\usepackage{slashed}
\usepackage[colorlinks,linkcolor=blue,anchorcolor=black,citecolor=blue]{hyperref}
\usepackage{multirow}
\usepackage{ulem}
\usepackage{comment}
\allowdisplaybreaks[4]

\begin{document}

\title{Twist-3 Contributions in Semi-Inclusive DIS in the Target Fragmentation Region}

\author{K.B. Chen}
\email{chenkaibao19@sdjzu.edu.cn}
\affiliation{School of Science, Shandong Jianzhu University, Jinan, Shandong 250101, China}

\author{J.P. Ma}
\email{majp@itp.ac.cn}
\affiliation{CAS Key Laboratory of Theoretical Physics, Institute of Theoretical Physics, P.O. Box 2735, Chinese Academy of Sciences, Beijing 100190, China}
\affiliation{School of Physical Sciences, University of Chinese Academy of Sciences, Beijing 100049, China}
\affiliation{School of Physics and Center for High-Energy Physics, Peking University, Beijing 100871, China}

\author{X.B. Tong}
\email{tongxuanbo1@gmail.com}
\affiliation{School of Science and Engineering, The Chinese University of Hong Kong, Shenzhen, Shenzhen, Guangdong, 518172, China}
\affiliation{University of Science and Technology of China, Hefei, Anhui, 230026, China}

\par\vskip10pt

\begin{abstract}
We present the complete results up to twist-3 for hadron production in the target fragmentation region  of  semi-inclusive deep inelastic scattering with a polarized lepton beam and polarized nucleon target.
The nonperturbative effects are factorized into fracture functions. 
The calculation up to twist-3 is nontrivial since one has to keep gauge invariance.
By applying collinear expansion, we show that the hadronic tensor can be expressed by gauge-invariant fracture functions.
We also present the results for the structure functions and azimuthal asymmetries.
\end{abstract}

\maketitle

\section{Introduction}
\label{sec:Introduction}

Semi-inclusive deep inelastic scattering (SIDIS) is an important process for hadronic physics. It provides 
a cleaner environment for detecting the inner structure of the initial hadron than inclusive processes in hadron-hadron collisions.  
The kinematic region of SIDIS can roughly be divided into two parts~(see e.g.,~\cite{Boglione:2022gpv,Boglione:2019nwk,Gonzalez-Hernandez:2018ipj,Boglione:2016bph,Mulders:2000jt} for more discussions). 
One is called the current fragmentation region~(CFR) where the observed hadron in the final state moves into the forward region of the virtual photon. 
Another one is the target fragmentation region~(TFR), where the measured hadron predominantly travels in the forward direction of the incoming target. 
Events in both regions can be used to comprehend the internal structure of hadrons and the properties of strong interactions.

So far, the bulk of research on SIDIS has focused on the CFR, where hadron production can be understood as the fragmentation of a parton emitted from the target and struck by the virtual photon. 
This allows us to investigate various parton distributions functions (PDFs)~\cite{Collins:1981uw,Ethier:2020way,Jimenez-Delgado:2013sma,Forte:2013wc} and fragmentation functions (FFs)~\cite{Collins:1981uw,Metz:2016swz,Chen:2023kqw} within the transverse-momentum-dependent~(TMD)~\cite{Collins:1981uk,Collins:2004nx,Ji:2004wu,Ji:2004xq,Collins:2011zzd,Rodini:2023plb,Boussarie:2023izj} or collinear factorization formalisms~\cite{Collins:1989gx,Meng:1991da,Daleo:2004pn,Kniehl:2004hf,Wang:2019bvb} at small or large hadron transverse momentum.
While there have been significant developments in recent years for physics in the CFR~\cite{Angeles-Martinez:2015sea,Scimemi:2019mlf,Boussarie:2023izj}, as well as elaborated and recent extractions of quark TMDs~\cite{Scimemi:2019cmh,Bacchetta:2019sam,Bacchetta:2022awv}, the physics in the TFR has received less attention.

The early analysis of the experimental data at HERA~\cite{ZEUS:1994mgw,H1:1995cha} indicates a surprisingly high number of events in the TFR and has stimulated the introduction of fracture functions~\cite{Trentadue:1993ka,Berera:1995fj,Grazzini:1997ih}. 
Physically, fracture functions describe the distributions of the struck parton inside the target when the remnant spectators fragment inclusively into the detected hadron. 
They encompass intricate initial final-state correlations and provide a unique perspective into the partonic dynamics and hadronization, complementing PDFs and FFs. 
Most of our current knowledge about fracture functions comes from the analysis of proton diffraction~(see e.g.,~\cite{Frankfurt:2022jns} for a recent review), where the final hadron coincides with the target proton, and the fracture function is conventionally called as diffractive PDF~\cite{Berera:1995fj}. 
Phenomenological fittings of diffractive PDFs from HERA data~\cite{H1:1994ahk,H1:1995cha,H1:1997bdi,H1:2006zyl,H1:2006uea,H1:2011jpo,H1:2012pbl,H1:2012xlc} have been conducted in~\cite{deFlorian:1998rj,Goharipour:2018yov,Khanpour:2019pzq,Maktoubian:2019ppi,Salajegheh:2022vyv,Salajegheh:2023jgi}. 
Fracture functions for other leading baryon production, such as neutron and $\Lambda$-hyperon, are also constrained with  parametrization assumptions in \cite{deFlorian:1997wi,Ceccopieri:2014rpa,Shoeibi:2017lrl,Shoeibi:2017zha,Mohsenabadi:2021bhs} and~\cite{Ceccopieri:2012rm,Ceccopieri:2015kya}, respectively. 
Fracture functions are also utilized in hadron collisions in \cite{Ceccopieri:2008fq,Ceccopieri:2010zu,Chai:2019ykk}.

From a theoretical point of view, most of the aforementioned investigations about TFR hadron production are based on the factorization at twist-2 in terms of collinear fracture functions.
This factorization has been proven to hold to all orders of $\alpha_s$~\cite{Collins:1997sr} and has been confirmed through the explicit calculations up to ${\cal O}(\alpha_s^2)$~\cite{Graudenz:1994dq,deFlorian:1995fd,Daleo:2003jf,Daleo:2003xg}. 
Initially, the fracture functions in the factorization included an integration over the final-hadron transverse momentum $P_{h\perp}$~\cite{Trentadue:1993ka}, however, without this integration the momentum transfer~\cite{Berera:1995fj,Grazzini:1997ih} and azimuthal-angle distribution~\cite{Anselmino:2011ss,Chen:2021vby} can be studied. To further probe the spin~\cite{Anselmino:2011ss,deFlorian:1995fd} and parton-transverse-momentum~\cite{Anselmino:2011ss,Ceccopieri:2005zz,Ceccopieri:2007ek} dependence of fracture functions, several observables and factorization assumptions are proposed in~\cite{Anselmino:2011bb,Anselmino:2011vkz,Yang:2020sos}. 
The CLAS Collaboration at JLab has recently reported the first measurement of these dependences~\cite{CLAS:2022sqt}. 
More detailed discussions on the factorization with TMD fracture functions and relevant evolution is presented in~\cite{Chai:2019ykk}.
Furthermore, recent investigations have also addressed the factorization properties of fracture functions in different kinematic regions~\cite{Chai:2019ykk,Chen:2021vby,Hatta:2022lzj}. The small-$x$ behavior of fracture functions is studied in~\cite{Hatta:2022lzj}.
In~\cite{Chen:2021vby} the large-$P_{h\perp}$ behavior is explored in detail, aiming to understand the transition of production mechanisms between the TFR and CFR in SIDIS.

Despite these progresses, the contributions of SIDIS in the TFR beyond the leading twist are still less investigated. The theoretical framework for a systematic study of the TFR beyond the leading twist needs to be established.
The importance of higher-twist effects in improving the description of the experimental data has already been emphasized by recent phenomenological studies of fracture functions~\cite{Golec-Biernat:2007mao,Maktoubian:2019ppi,Salajegheh:2022vyv}.

Moreover, it is known that the absence of higher-twist effects results in the loss of predictions for fourteen SIDIS structure functions in the TFR at the tree level. At the leading twist, only four structure functions are nonzero~\cite{Anselmino:2011ss} for the case of unpolarized hadron production and a spin-1/2 target~\cite{Diehl:2005pc,Bacchetta:2006tn}. 
These higher-twist contributions are responsible for various intriguing azimuthal and spin asymmetries. 
Especially, some of these asymmetries are already within the reach of the ongoing experimental program by CLAS12~\cite{Burkert:2020akg} at JLab due to the availability of a longitudinally polarized target~\cite{Arrington:2021alx,Accardi:2023chb}. 
For instance, a preliminary investigation utilizing CLAS12 data has revealed the significance of the beam-spin asymmetry in the TFR, suggesting that its sign and magnitude could serve as a novel indicator for tracking the transition between the TFR and the CFR~(Sec. 5.3 in~\cite{Accardi:2023chb}). 
Furthermore, the potential JLab@22GeV program~\cite{Accardi:2023chb} and the planned electron-ion colliders in the USA~\cite{Boer:2011fh,Accardi:2012qut,AbdulKhalek:2021gbh,AbdulKhalek:2022hcn,Abir:2023fpo} and China~\cite{Anderle:2021wcy} are poised to provide additional exciting opportunities for exploring new frontiers in TFR physics. 
Given these experimental progresses, it is important to undertake an evaluation of the higher-twist contributions of SIDIS in the TFR.

The objective of this paper is to present a first analysis of twist-$3$ contributions to SIDIS in the TFR within the framework of collinear factorization at the tree level of quantum chromodynamics~(QCD) perturbation theory. 
Our focus is on the scenario where the target is spin-1/2 and the polarization of the final hadron is unobserved. 
The framework can be easily extended to the case of a spin-1 target.
By employing the collinear expansion technique~\cite{Ellis:1982wd,Ellis:1982cd,Qiu:1988dn,Qiu:1990xxa,Qiu:1990xy,Liang:2006wp,Song:2010pf,Wei:2014pma,Yang:2017sxz}, we demonstrate that the hadronic tensor of SIDIS in the TFR can be expressed in terms of three distinct types of twist-3 collinear fracture functions. 
We discuss the classification of these fracture functions and show that 
they are not independent due to the constraints imposed by the QCD equation of motion~(EOM). 
With the EOM, the twist-3 contributions can be expressed with two-parton fracture functions at the considered order.  Our findings also have significant phenomenological implications.

The rest of this paper is organized as follows.
In Sec.~\ref{sec:Kinematics}, we discuss the kinematics for the polarized SIDIS in TFR and present the general form for the cross section in terms of the structure functions.
In Sec.~\ref{sec:HadronicTensor}, we present detailed calculations of the hadronic tensor up to twist-3.
In Sec.~\ref{sec:FinalResults}, we give the final results for the structure functions and azimuthal or spin asymmetries expressed by fracture functions.
A short summary is given in Sec.~\ref{sec:Summary}.

\section{Kinematics and Structure Functions of SIDIS in the TFR}
Through out this paper, we use the light-cone coordinate system, in which a vector $a^\mu$ is expressed as $a^\mu = (a^+,a^-, \vec a_\perp) = \bigl((a^0+a^3)/{\sqrt{2}}, (a^0-a^3)/{\sqrt{2}}, a^1, a^2 \bigr)$.
With the light cone vectors $n^\mu = (0,1,0,0)$ and $\bar n^\mu = (1,0,0,0)$, the transverse metric is defined as $g_\perp^{\mu\nu} = g^{\mu\nu} - \bar n^\mu n^\nu - \bar n^\nu n^\mu$, and the transverse antisymmetric tensor is given as $\varepsilon_\perp^{\mu\nu} = \varepsilon^{\mu\nu\alpha\beta} \bar n_\alpha n_\beta$ with $\varepsilon_\perp^{12} = 1$. We also use the notation $\tilde a_\perp^\mu \equiv \varepsilon_\perp^{\mu\nu} a_{\perp\nu}$.

\label{sec:Kinematics}
We consider the SIDIS process with a polarized electron beam and nucleon target as follows:
\begin{equation} 
e(l,\lambda_e) + h_A (P,S) \to e(l^\prime) + h(P_h) + X, 
\end{equation} 
where $l$, $l^\prime$, $P$, and $P_h$ are the 4-momenta of the incident, the outgoing electron, the nucleon target and the detected final-state hadron, respectively.
At the leading order of quantum electrodynamics, there is an exchange of one virtual photon between the electron and the nucleon.
The momentum of the virtual photon is given by $q=l-l^\prime$.
The helicity of the electron is denoted by $\lambda_e$, and $S$ is the polarization vector of the nucleon.
We consider the production of a spin-0 or unpolarized final-state hadron $h$. 
 The Lorentz invariant variables of SIDIS are conventionally defined by
\begin{align}
 Q^2 = -q^2,~ x_B = \frac{Q^2}{2 P\cdot q},~  y=\frac{ P\cdot q}{P\cdot k_e},~ z_h=\frac{P\cdot P_h}{P\cdot q}. 
\end{align}
We are interested in the TFR, where $P_h$ is almost collinear with $P$ and $z_h\ll 1$. As discussed in ~\cite{Graudenz:1994dq,Anselmino:2011ss}, $z_h$ is not convenient for us to describe the hadron production in TFR, because one can not differentiate the scenario of TFR considered here from the soft-hadron production. Instead, we will use~\cite{Anselmino:2011ss,Boglione:2019nwk}
\begin{align}
\xi_h = \frac{P_h \cdot q}{P \cdot q}.
\end{align}
\begin{figure}[htb]
\centering
\includegraphics[width=0.35\textwidth]{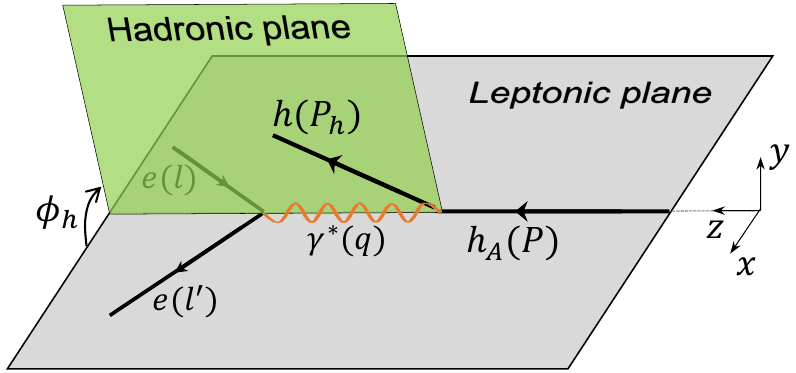}
\caption{The kinematics for SIDIS in the TFR}
\label{fig:kinematics}
\end{figure}

We work in the reference frame shown in Fig.~\ref{fig:kinematics},
where the nucleon $h_A$ moves along the $+z$-direction and the virtual photon moves in the $-z$-direction. 
In this frame, the momenta of the particles are given by
\begin{align} 
& P^\mu \approx ( P^+,0,0,0), \\
& P_h^\mu = (P_h^+, P_h^-, \vec P_{h\perp}), \\
& l^\mu = \Big(\frac{1-y}{y}x_B P^+,~ \frac{Q^2}{2x_B y P^+},~ \frac{Q\sqrt{1-y}}{y},~0\Big), \\
& q^\mu =\Big(-x_B P^+,~ \frac{Q^2}{2x_BP^+}, 0,0\Big).
\end{align}
For the case that the produced hadron $h$ has small transverse momentum and in the TFR, we have $P_h^+ \gg |\vec P_{h\perp}| \gg P_h^-$ and $\xi_h \approx P_h^+/P^+$,
which specifies the longitudinal momentum fraction of the nucleon taken by the final-state hadron $h$.
The polarization vector of the nucleon with mass $M$ can be decomposed by
\begin{align}
S^\mu = S_L \frac{P^+}{M} \bar n^\mu + S_\perp^\mu - S_L \frac{M}{2P^+} n^\mu,
\end{align}
where $S_L$ is the longitudinal polarization of the nucleon and $S_\perp^\mu = (0,0,\vec S_\perp)$ the transverse polarization vector.
 
The incoming and outgoing electron span the lepton plane. 
We define the azimuthal angle $\phi_h$ for $\vec P_{h\perp}$ with respect to the lepton plane, and $\phi_S$ is that for $\vec S_\perp$.
The azimuthal angle of the outgoing lepton around the lepton beam with respect to the spin vector is denoted by $\psi$.
In the kinematic region of SIDIS with large $Q^2$, one has $\psi\approx \phi_S$~\cite{Diehl:2005pc}.
With these specifications, the differential cross section is given by
\begin{align} 
\frac{ d\sigma}{dx_B dy d\xi_h d\psi d^2 P_{h\perp} } = \frac{\alpha^2 y}{4 \xi_h Q^4}   L_{\mu\nu} (l,\lambda_e,l^\prime) W^{\mu\nu}(q,P,S,P_h),
\label{eq:CrossSection}
\end{align}
where $\alpha$ is the fine structure constant.
The leptonic tensor is
\begin{align}
L^{\mu\nu}(l,\lambda_e,l^\prime)=
2 (l^\mu l^{\prime\nu} + l^\nu l^{\prime\mu} - l\cdot l^\prime g^{\mu\nu}) +  2i \lambda_e \epsilon^{\mu \nu\rho\sigma} l_\rho l^\prime_\sigma.
\label{eq:LeptonicTensor}
\end{align}
The hadronic tensor is defined by
\begin{align} 
W^{\mu\nu}(q,P,S,P_h) = \sum_X \int \frac {d^4 x}{(2\pi)^4} e^{iq\cdot x} \langle S;h_A \vert J^\mu (x) \vert h X\rangle \langle X h \vert J^\nu (0) \vert h_A;S \rangle,
\label{eq:Wuv}
\end{align}
where $J^\mu(x) = e_q \bar \psi(x) \gamma^\mu \psi(x)$ is the electromagnetic current.
A summation over quark favors is implicit in Eq.~(\ref{eq:Wuv}).

In general, the hadronic tensor can be decomposed into a sum of basic Lorentz tensors constructed by the kinematic variables of the process. 
After contracting with the leptonic tensor, one can get the differential cross section in terms of the structure functions. 
It has been shown that the differential cross section of SIDIS at small transverse momentum with the polarized lepton beam and nucleon target is described by eighteen structure functions~\cite{Bacchetta:2006tn}. 
We have the same number of structure functions for SIDIS in the TFR, and the general form of the differential cross section can be expressed as
\begin{align}
& \frac{ d\sigma}{d x_B d y d \xi_h  d\psi d^2 P_{h\perp} } =\frac{\alpha^2}{x_B y Q^2} \Bigl\{ A(y) F_{UU,T} + E(y) F_{UU,L} + B(y) F_{UU}^{\cos\phi_h} \cos\phi_h + E(y) F_{UU}^{\cos2\phi_h} \cos2\phi_h \nonumber\\
& + \lambda_e D(y) F_{LU}^{\sin\phi_h} \sin\phi_h + S_L \Bigl[ B(y) F_{UL}^{\sin\phi_h} \sin\phi_h + E(y) F_{UL}^{\sin2\phi_h} \sin2\phi_h \Bigr] + \lambda_e S_L \Bigl[ C(y) F_{LL} + D(y) F_{LL}^{\cos\phi_h} \cos\phi_h \Bigr] \nonumber\\
& + |\vec S_\perp| \Bigl[ \bigl( A(y) F_{UT,T}^{\sin(\phi_h-\phi_S)} + E(y) F_{UT,L}^{\sin(\phi_h-\phi_S)} \bigr) \sin(\phi_h-\phi_S) + E(y) F_{UT}^{\sin(\phi_h+\phi_S)} \sin(\phi_h+\phi_S) \nonumber\\
& \qquad + B(y) F_{UT}^{\sin\phi_S} \sin\phi_S + B(y) F_{UT}^{\sin(2\phi_h-\phi_S)} \sin(2\phi_h-\phi_S) + E(y) F_{UT}^{\sin(3\phi_h-\phi_S)} \sin(3\phi_h-\phi_S) \Bigr] \nonumber\\
& + \lambda_e |\vec S_\perp| \Bigl[ D(y) F_{LT}^{\cos\phi_S} \cos\phi_S + C(y) F_{LT}^{\cos(\phi_h-\phi_S)} \cos(\phi_h-\phi_S) + D(y) F_{LT}^{\cos(2\phi_h-\phi_S)} \cos(2\phi_h-\phi_S) \Bigr] \Bigr\}.
\label{eq:SFs-SIDIS}
\end{align}
Here we have defined several functions of $y$ for convenience, i.e.,
\begin{align}
& A(y) = y^2-2y+2, \nonumber\\
& B(y) = 2(2-y)\sqrt{1-y}, \nonumber\\
& C(y) = y(2-y), \nonumber\\
& D(y) = 2y\sqrt{1-y}, \nonumber\\
& E(y) = 2(1-y).
\end{align}
All the structure functions in Eq.~(\ref{eq:SFs-SIDIS}) are scalar functions depending on $x_B$, $\xi_h$, $Q^2$ and $\vec P_{h\perp}^2$.
The first and second subscripts of the structure functions denote the polarization of the electron and the nucleon, respectively.
The third subscript, if any, specifies the polarization of the virtual photon. 
Note that the normalization of the structure functions adopted here is different from that in~\cite{Bacchetta:2006tn} by a Jacobian since we have used $\xi_h$ instead of $z_h$.

\section{The hadronic tensor results up to twist-3}
\label{sec:HadronicTensor}
\subsection{Collinear expansion for the hadronic tensor}
\begin{figure}[htb]
\centering
\includegraphics[width=0.75\textwidth]{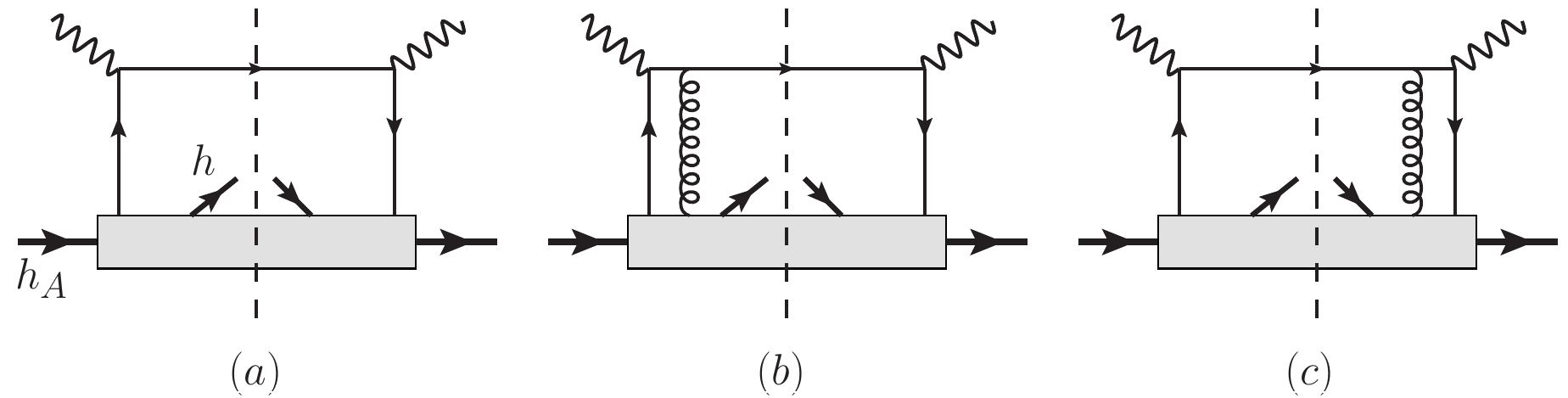}
\caption{Diagrams for the hadronic tensor in TFR at tree level.}
\label{fig:Wuv}
\end{figure}
Now we perform the collinear expansion for the hadronic tensor in Eq.~(\ref{eq:Wuv}) up to twist-3. 
At the tree level of QCD perturbation theory, the hadronic tensor in the TFR can be represented by the diagrams in Fig.~\ref{fig:Wuv}. 
The gray boxes represent the parton correlation matrices with a hadron $h$ identified in the final state, which we call fracture matrices in the following. 
The contributions for each diagram in Fig.~\ref{fig:Wuv} are
\begin{align}
W^{\mu\nu} \Big\vert_{2a} =& \int \frac{d^3 k}{(2\pi)^3} 
 \left[ \left(\gamma^\mu  (\slashed k +\slashed q) \gamma^\nu\right)_{ij} 2\pi \delta \bigl((k+q)^2\bigr) \right]  \sum_X\int \frac{d^3 \eta}{(2\pi)^4} e^{-i k \cdot \eta} \langle h_A|\bar \psi_i(\eta)|hX \rangle \langle Xh|  \psi_j(0) |h_A \rangle \label{eq:Wuv-a}, \\
W^{\mu\nu} \Big\vert_{2b} =& \int \frac{d^3 k_1 d^3 k_2}{(2\pi)^6}  \left[ \left( \gamma^\mu (\slashed k_1 +\slashed q) \gamma_\alpha \frac{i(\slashed k_2 + \slashed q)}{(k_2+q)^2 + i\epsilon} \gamma^\nu \right)_{ij} 2\pi \delta \bigl((k_1+q)^2\bigr) \right]  \nonumber\\
 &\times (-ig_s) \sum_X\int \frac{d^3 \eta d^3 \eta_1}{(2\pi)^4}  e^{-i k_1 \cdot \eta} e^{i(k_1-k_2)\cdot \eta_1}
 \langle h_A|\bar \psi_i(\eta)|hX \rangle   \langle Xh| G^{\alpha}(\eta_1) \psi_j(0) |h_A \rangle,\label{eq:Wuv-b} \\
W^{\mu\nu} \Big\vert_{2c} =& \int \frac{d^3 k_1 d^3 k_2}{(2\pi)^6} \left[ \left( \gamma^\mu \frac{i(\slashed k_1 + \slashed q)}{(k_1+q)^2 - i\epsilon} \gamma_\alpha (\slashed k_2 +\slashed q)   \gamma^\nu \right)_{ij}  2\pi \delta \bigl((k_2+q)^2\bigr) \right] \nonumber\\
& \times (-ig_s) \sum_X \int \frac{d^3 \eta d^3 \eta_1}{(2\pi)^4} e^{-i k_1 \cdot \eta} e^{i(k_1-k_2)\cdot \eta_1}
\langle h_A|\bar \psi_i(\eta) G^{\alpha}(\eta_1) |hX \rangle   \langle Xh| \psi_j(0) |h_A \rangle, \label{eq:Wuv-c}
\end{align}
where $ij$ are the Dirac and color indices. The summation over quark flavors $\sum_q e_q^2$ is implied in the expressions. The integration variables take the following forms:
\begin{align}
& k^\mu = (k^+, 0, \vec k_\perp), \quad k_1^\mu = (k_1^+, 0, \vec k_{1\perp}), \quad k_2^\mu = (k_2^+, 0, \vec k_{2\perp}), \\
& \eta^\mu = (0, \eta^-, \vec \eta_\perp), \quad \eta_1^\mu = (0, \eta_1^-, \vec \eta_{1\perp}), \quad \eta_2^\mu = (0, \eta_2^-, \vec \eta_{2\perp}).
\end{align}
$k$ is the momentum carried by the quark line leaving the box of Fig.~\ref{fig:Wuv}(a).
$k_1$, $k_2$ are the momenta carried by the quark lines flowing into and out of the boxes of Figs.~\ref{fig:Wuv}(b) or \ref{fig:Wuv}(c).  These momenta follow the collinear scaling, e.g., $k^\mu\sim Q(1,\lambda^2,\lambda)$ with $\lambda=\Lambda_{\text{QCD}}/Q$.
To obtain the contributions up to twist-3, one has to expand the contributions in Figs.~\ref{fig:Wuv}(a)-\ref{fig:Wuv}(c) in powers of $\lambda$ up to ${\cal O}(\lambda)$. 
Here we have already neglected the minus components of $k$, $k_1$ and $k_2$ in $[\cdots]$ of Eqs.~(\ref{eq:Wuv-a})-(\ref{eq:Wuv-c}), since these components only yield the corrections beyond twist-3.

For $W^{\mu\nu}|_{2a}$, if we further neglect the quark transverse momentum and take $k^\mu \approx (k^+,0,\vec 0_\perp)$ in $[\cdots]$ of Eq.~(\ref{eq:Wuv-a}), one can obtain the contribution with the collinear fracture matrix involving a nonlocal operator of quark and antiquark fields. 
To obtain a gauge-invariant form, we should sum over the contributions from the $G^+$-gluon exchange in Fig.~\ref{fig:Wuv}(b) and Fig.~\ref{fig:Wuv}(c)  as well as those with the exchange of any number of $G^+$-gluons. 
Here the gluon field $G^\mu$ scales like $(1,\lambda^2,\lambda)$, and hence the $G^+$-gluon does not induce any power suppression. 
After this summation, we can obtain the following gauge-invariant contribution
\begin{align}
W^{\mu\nu} \Big\vert_{\text{q}} =& (\gamma^\mu \gamma^+ \gamma^\nu)_{ij} \sum_X\int \frac{d\eta^-}{2(2\pi)^4} e^{-ix_BP^+\eta^-} \langle h_A|\bar \psi_i(\eta^-) {\cal L}_n^{\dagger}(\eta^-) |hX \rangle  \langle Xh| {\cal L}_n(0) \psi_j(0) |h_A \rangle, \label{eq:Wuv-2p}
\end{align}
where the gauge link is defined as
\begin{align}
{\cal L}_n (x) = {\cal P} \exp \biggr \{ - i g_s \int_0^\infty d\lambda ~G^{+}(\lambda n +x)  \biggr \}.
\end{align}
The above contribution yields the gauge-invariant collinear quark fracture matrix. As we will present later in Sec.~\ref{sec:par}, by parametrization of this matrix up to ${\cal O}(\lambda)$, one can obtain the hadronic tensor in terms of twist-2 and twist-3 quark collinear fracture functions. 

To derive the other twist-3 contributions from $W^{\mu\nu}\vert_{2a}$, we need to take into account the $k_\perp$-dependence in $[\cdots]$ of Eq.~(\ref{eq:Wuv-a}) by the collinear expansion to ${\cal O}(\lambda)$. We notice that
 there is no contribution from the partial derivatives acting on the delta function in the expansion since $\partial\delta\bigl((\hat k+q)^2\bigr)/\partial k_{\perp}^\alpha \propto  q_{\perp\alpha} = 0$. After this expansion, we get the contribution from the fracture matrix with the transverse partial derivative acting on the (anti)quark fields. Again, after the combination with the relevant gauge-link contributions from $W^{\mu\nu}\vert_{2b+2c}$, we obtain, up to ${\cal O}(g_s)$,
\begin{align}
    W^{\mu\nu} \Big\vert_{\partial} =& \frac{-i}{2q^-} (\gamma^\mu \gamma^+\gamma_{\perp\alpha} \gamma^- \gamma^\nu)_{ij} \sum_X\int \frac{d \eta^-}{2(2\pi)^4}  e^{-i x_B P^+ \eta^-}  \langle h_A| \bar \psi_i(\eta^-){\cal L}_n^{\dagger}(\eta^-)|hX \rangle \nonumber\\
& \qquad \times \langle Xh| \partial_\perp^\alpha({\cal L}_n\psi_j)(0) |h_A \rangle
+ (\mu \leftrightarrow \nu)^*~,\label{eq:Wuv-partial}
\end{align}
where $(\mu \leftrightarrow \nu)^*$ stands for exchanging $\mu\nu$ indices and taking complex conjugate of the first term. Due to the presence of the transverse derivative, the leading contribution of the fracture matrix in Eq.~(\ref{eq:Wuv-partial}) is at twist-3.
\par 
In addition, after subtracting the gauge-link contributions to Eqs.~(\ref{eq:Wuv-2p}) and (\ref{eq:Wuv-partial}) from the collinear expansion of $W^{\mu\nu}\vert_{2b+2c}$, we find the remaining part can be expressed by the fracture matrix with the gluon field-strength tensor $g_s F^{+\alpha} = g_s[\partial^+G_\perp^\alpha - \partial_\perp^\alpha G^+]+{\cal O}(g_s^2)$. This gives another contribution that starts from twist-3, which up to ${\cal O}(g_s)$ can be summarized as
\begin{align} 
W^{\mu\nu} \Big\vert_{F} =& \frac{-i}{2q^-} (\gamma^\mu \gamma^+\gamma_{\perp\alpha} \gamma^- \gamma^\nu)_{ij} \int dx_2 \Bigl[ {\rm P} \frac{1}{x_2 - x_B} - i\pi \delta(x_2-x_B) \Bigr] \nonumber\\
&\times \sum_X\int \frac{d\eta^- d\eta_1^-}{4\pi(2\pi)^4} e^{-ix_B P^+ \eta^- - i(x_2-x_B) P^+ \eta_1^-} \langle h_A | \bar \psi_i(\eta^-) |hX\rangle \langle Xh | g_s F^{+\alpha}(\eta_1^-) \psi_j(0) |h_A \rangle + (\mu \leftrightarrow \nu)^*~.\label{eq:Wuv-F}~
\end{align}
Here $\rm P$ in $[\cdots]$ of Eq.~(\ref{eq:Wuv-F}) stands for the principle-value prescription.
The $\delta$-function term in Eq.~(\ref{eq:Wuv-F}) comes from the absorptive part of the quark propagator that connects the electromagnetic current to the quark-gluon vertex in Figs.~\ref{fig:Wuv}(b) and \ref{fig:Wuv}(c).
In this term, the gluon has zero momentum and generates the so-called soft-gluon-pole contributions, see e.g.,~\cite{Chen:2015uqa} and references therein.

The total contribution of the hadronic tensor is given by the sum of the results in Eqs.~(\ref{eq:Wuv-2p}), (\ref{eq:Wuv-partial}) and (\ref{eq:Wuv-F}). The following gauge-invariant collinear fracture matrices are relevant:
\begin{align}
& {\cal M}_{ij}(x) =\int \frac{d\eta^-}{2\xi_h(2\pi)^4} e^{-ixP^+\eta^-} \sum_X \langle h_A|\bar \psi_j(\eta^-) {\cal L}_n^{\dagger}(\eta^-) |hX \rangle  \langle Xh| {\cal L}_n(0) \psi_i(0) |h_A \rangle, \label{eq:M2p} \\
& {\cal M}_{\partial,ij}^\alpha(x) = \int \frac{d \eta^-}{2\xi_h(2\pi)^4}  e^{-i x P^+ \eta^-}  \sum_X\langle h_A| \bar \psi_j(\eta^-){\cal L}_n^{\dagger}(\eta^-)|hX \rangle \langle Xh| \partial_\perp^\alpha({\cal L}_n\psi_i)(0) |h_A \rangle, \label{eq:Mpartial} \\
& {\cal M}_{F,ij}^\alpha(x_1,x_2) = \int \frac{d\eta^- d\eta_1^-}{4\pi\xi_h(2\pi)^4} e^{-ix_1 P^+ \eta^- - i(x_2-x_1) P^+ \eta_1^-}\sum_X \langle h_A | \bar \psi_j(\eta^-) |hX\rangle \langle Xh | g_s F^{+\alpha}(\eta_1^-)  \psi_i(0) |h_A \rangle. \label{eq:MF}
\end{align}
Here we have suppressed the gauge links for brevity in Eq.~(\ref{eq:MF}). Besides the partonic momentum fractions, the fracture matrices also depend on the momentum of the observed hadron~($\xi_h$, $P_{h\perp}$) and the spin vector of the target, which are not shown explicitly in the arguments. With the above notations, the hadronic tensor can be written as
\begin{align}
W^{\mu\nu} &= \xi_h(\gamma^\mu \gamma^+\gamma^\nu)_{ij} {\cal M}_{ji}(x_B) 
+ \left[ \frac{-i\xi_h}{2q^-} (\gamma^\mu \gamma^+\gamma_{\perp\alpha} \gamma^- \gamma^\nu)_{ij} {\cal M}_{\partial,ji}^\alpha(x_B) + (\mu \leftrightarrow \nu)^* \right] \nonumber\\
&+ \left\{ \frac{-i\xi_h}{2q^-} (\gamma^\mu \gamma^+\gamma_{\perp\alpha} \gamma^- \gamma^\nu)_{ij} \int dx_2 \Bigl[ {\rm P} \frac{1}{x_2 - x_B} - i\pi \delta(x_2-x_B) \Bigr] {\cal M}_{F,ji}^\alpha(x_B,x_2) + (\mu \leftrightarrow \nu)^* \right\}. \label{eq:Wuv-total}
\end{align}
It is obvious that only the chirality-even parts of the fracture matrices will contribute to the hadronic tensor.
We discuss the parametrization of these fracture matrices in the next subsection.

\subsection{Parametrization of the fracture matrices and the hadronic tensor in terms of fracture functions \label{sec:par}}

The collinear fracture matrices in Eqs.~(\ref{eq:M2p})-(\ref{eq:MF}) can be decomposed using Dirac $\Gamma$-matrices. From the constraints of parity invariance, the fracture matrices can be generally parametrized as follows:
\begin{align}
{\cal M}_{ij}(x) &=\frac{(\gamma_\rho)_{ij} }{2N_c}\Bigl[ \bar n^\rho \Bigl( u_1 - \frac{P_{h\perp} \cdot \tilde S_\perp}{M} u_{1T}^{h} \Bigr) + \frac{1}{P^+}\Bigl( P_{h\perp}^{\rho} u^{h}  - M \tilde S_{\perp}^{\rho} u_T  - S_L \tilde P_{h\perp}^{\rho} u_L^{h} - \frac{P_{h\perp}^{\langle\rho} P_{h\perp}^{\beta\rangle}}{M} \tilde S_{\perp\beta} u_T^{h}  \Bigr)\Bigr] \nonumber\\
&- \frac{(\gamma_\rho\gamma_5)_{ij} }{2N_c}\Bigl[ \bar n^\rho \Bigl( S_L l_{1L}  - \frac{P_{h\perp} \cdot S_\perp}{M} l_{1T}^{h}  \Bigr) + \frac{1}{P^+}\Bigl( \tilde P_{h\perp}^\rho l^{h}  + M S_{\perp}^\rho l_T  + S_L P_{h\perp}^\rho l_L^{h}  - \frac{P_{h\perp}^{\langle\rho} P_{h\perp}^{\beta\rangle}}{M} S_{\perp\beta} l_T^{h}  \Bigr) \Bigr] + \cdots, \label{eq:FrF-u}\\
{\cal M}_{\partial,ij}^\alpha(x) &= \frac{(\gamma^-)_{ij}}{2N_c} i\Bigl( -P_{h\perp}^\alpha u_\partial^{h}  + M\tilde S_\perp^\alpha u_{\partial T}  + S_L \tilde P_{h\perp}^\alpha u_{\partial L}^{h}  + \frac{P_{h\perp}^{\langle\alpha} P_{h\perp}^{\beta\rangle}}{M} \tilde S_{\perp\beta} u_{\partial T}^{h}  \Bigr) \nonumber\\
&+\frac{(\gamma^-\gamma_5)_{ij}}{2N_c} i \Bigl( \tilde P_{h\perp}^\alpha l_\partial^{h}  + M S_{\perp}^\alpha l_{\partial T}  + S_L P_{h\perp}^\alpha l_{\partial L}^{h}  - \frac{P_{h\perp}^{\langle\alpha} P_{h\perp}^{\beta\rangle}}{M} S_{\perp\beta} l_{\partial T}^{h}  \Bigl) + \cdots, \label{eq:FrF-upartial}\\
{\cal M}_{F,ij}^\alpha(x_1,x_2) &= \frac{(\gamma^-)_{ij}}{2N_c} \Bigl( P_{h\perp}^{\alpha} w^{h} - M \tilde S_{\perp}^{\alpha} w_{T} - S_L \tilde P_{h\perp}^{\alpha} w_{L}^{h} - \frac{P_{h\perp}^{\langle\alpha} P_{h\perp}^{\beta\rangle}}{M} \tilde S_{\perp\beta} w_{T}^{h} \Bigr) \nonumber\\
& -  \frac{(\gamma^-\gamma_5)_{ij}}{2N_c}  i \Bigl( \tilde P_{h\perp}^\alpha v^{h} + M S_{\perp}^\alpha v_{ T} + S_L P_{h\perp}^\alpha v_{ L}^{h} - \frac{P_{h\perp}^{\langle\alpha} P_{h\perp}^{\beta\rangle}}{M} S_{\perp\beta} v_{T}^{h} \Bigr) + \cdots, \label{eq:FrF-uf}
\end{align}
where $\cdots$ denote the contributions beyond twist-3 or the chirality-odd parts. 
In the above, we have used the shorthand notations $P_{h\perp}^{\langle\alpha} P_{h\perp}^{\beta\rangle} \equiv P_{h\perp}^{\alpha} P_{h\perp}^{\beta} + g_{\perp}^{\alpha\beta}\vec P_{h\perp}^2/2$ for simplicity.
As pointed out in~\cite{Anselmino:2011ss}, the fracture matrix is not constrained by time reversal invariance, as it identifies a hadron in the out state. 
Additionally, we note that the collinear fracture matrices formally share similar parametrization forms with those of the conventional TMD PDFs~(see e.g.,~\cite{Wei:2016far}).

The functions $u$'s and $l$'s in Eqs.~(\ref{eq:FrF-u}) and (\ref{eq:FrF-upartial}) are quark collinear fracture functions, 
they are functions of $x$,  $\xi_h$ and $\vec P_{h\perp}^2$.  $w$'s and $v$'s in Eq.~(\ref{eq:FrF-uf}) are quark-gluon collinear fracture functions, they depend on $x_1$ and $x_2$ besides of $\xi_h$ and $\vec P_{h\perp}^2$.
We have suppressed all these arguments for simplicity.  
From hermiticity, the fracture functions defined in Eqs.~(\ref{eq:FrF-u}) and (\ref{eq:FrF-upartial}) are real, while those in Eq.~(\ref{eq:FrF-uf}) are complex in general.

The naming rules for these fracture functions we have used are as follows:
Four fracture functions in Eq.~(\ref{eq:FrF-u}) with ``1" in the subscript are of twist-2.
The remaining is of twist-3.
The ``$\partial$" in the subscript denote that the fracture functions are defined via the fracture matrix with the partial derivative operator.
The ``$L$" or ``$T$" in the subscript denotes the dependence on the longitudinal or transverse polarization of the nucleon.
The superscript ``$h$" denotes the explicit dependence on the transverse momentum of the final-state hadron $h$ in the decomposition of the matrix elements.
We note that the TMD quark fracture functions at twist-2 have been classified for a polarized nucleon target in~\cite{Anselmino:2011ss,Anselmino:2011bb}.
After integrating over the transverse momentum of the parton, they are equivalent to the twist-2 collinear quark fracture functions defined in Eq.~(\ref{eq:FrF-u}).

We further note that the twist-3 fracture functions defined above are not independent of each other. 
From the QCD equation of motion $i\gamma \cdot D \psi = 0$, one can show that their relations can be written in a unified form as follows:
\begin{align}
x[u_S^K(x) + il_S^K(x)] = u_{\partial S}^K(x) + il_{\partial S}^K(x) + i\int dy\Bigl[ {\rm P} \frac{1}{y-x} - i\pi\delta(y-x) \Bigr] [w_{S}^K(x,y) - v_{S}^K(x,y)], \label{eq:EMO}
\end{align}
where $(S,~K) = (\text{null},~ h)$, $(L,~ h)$, $(T,~ \text{null})$, or $(T,~ h)$.
I.e., we have four sets of relations in the unified form of Eq.~(\ref{eq:EMO}).
With these relations, we find that the hadronic tensor in Eq.~(\ref{eq:Wuv-total}) can be expressed only with the fracture functions defined via ${\cal M}_{ij}$ in Eq.~(\ref{eq:FrF-u}). 
We obtain
\begin{align}
W^{\mu\nu} 
& = -2g_\perp^{\mu\nu} \Bigl(u_1 - \frac{P_{h\perp} \cdot \tilde S_\perp}{M} u_{1T}^{h} \Bigr) 
+ 2i\varepsilon_\perp^{\mu\nu} \Bigl( S_L l_{1L} - \frac{P_{h\perp} \cdot S_\perp}{M} l_{1T}^{h} \Bigr) \nonumber\\
& + \frac{2}{P\cdot q} P_{h\perp}^{\{\mu} \bar q^{\nu\}} \Bigl( u^{h} - \frac{P_{h\perp} \cdot \tilde S_\perp}{M} u_{T}^{h} \Bigr) 
+ \frac{2i}{P\cdot q} P_{h\perp}^{[\mu} \bar q^{\nu]} \Bigl( l^{h} - \frac{P_{h\perp} \cdot \tilde S_\perp}{M} l_{T}^{h} \Bigr) - \frac{2M}{P\cdot q} \tilde S_\perp^{\{\mu} \bar q^{\nu\}} \Bigl( u_T - \frac{\vec P_{h\perp}^2}{2M^2} u_T^{h} \Bigr) \nonumber\\
& - \frac{2iM}{P\cdot q} \tilde S_\perp^{[\mu} \bar q^{\nu]} \Bigl( l_T - \frac{\vec P_{h\perp}^2}{2M^2} l_T^{h} \Bigr) - \frac{2S_L}{P\cdot q} \tilde P_{h\perp}^{\{\mu} \bar q^{\nu\}} u_L^{h} - \frac{2iS_L}{P\cdot q} \tilde P_{h\perp}^{[\mu} \bar q^{\nu]} l_L^{h},
\label{eq:HadronicTensor}
\end{align}
where $A^{\{\mu}B^{\nu\}} \equiv A^\mu B^\nu + A^\nu B^\mu$ and $A^{[\mu}B^{\nu]} \equiv A^\mu B^\nu - A^\nu B^\mu$.
We have also used the shorthand notation $\bar q^\mu \equiv q^\mu + 2x_B P^+ \bar n^\mu$.
The first line in Eq.~(\ref{eq:HadronicTensor}) is of twist-2 contributions, and the remains are of twist-3 contributions. 
Because $q^\mu$ has only longitudinal components and also $q\cdot \bar q = 0$,
we see explicitly that the hadronic tensor of Eq.~(\ref{eq:HadronicTensor}) satisfies the $U(1)$-gauge invariance or the current conservation, i.e., $q_\mu W^{\mu\nu} = q_\nu W^{\mu\nu} = 0$.

\section{The results of structure functions and azimuthal or spin asymmetries}
\label{sec:FinalResults}
\subsection{The results of structure functions}
Substituting the hadronic tensor result of Eq.~(\ref{eq:HadronicTensor}) into Eq.~(\ref{eq:CrossSection}),  we obtain the differential cross section.  
Comparing with the cross section expressed by structure functions in Eq.~(\ref{eq:SFs-SIDIS}), we obtain the results of structure functions in terms of the collinear fracture functions.
Four structure functions are at twist-2, which are expressed in terms of the four twist-2 fracture functions, i.e.,
\begin{align}
& F_{UU,T} = x_B u_1, \qquad F_{UT,T}^{\sin(\phi_h-\phi_S)} = \frac{|\vec  P_{h\perp}|}{M} x_B u_{1T}^{h}, \label{eq:FU}\\
& F_{LL} = x_B l_{1L}, \qquad F_{LT}^{\cos(\phi_h-\phi_S)} = \frac{|\vec  P_{h\perp}|}{M} x_B l_{1T}^{h}. \label{eq:FL}
\end{align}
The summation over quark flavors, i.e., $\sum_q e_q^2\cdots$, is implicit on the right-hand side of the equations.
This twist-2 result has been obtained in~\cite{Anselmino:2011bb}.
There are eight structure functions that have contributions starting from twist-3.
They are expressed with eight different twist-3 fracture functions, i.e.,
\begin{align}
& F_{UU}^{\cos\phi_h} = - \frac{2|\vec  P_{h\perp}|}{Q} x_B^2 u^{h}, \qquad F_{LU}^{\sin\phi_h} = \frac{2|\vec  P_{h\perp}|}{Q} x_B^2 l^{h}, \label{eq:Ftwist3-1} \\
& F_{UL}^{\sin\phi_h} = - \frac{2|\vec  P_{h\perp}|}{Q} x_B^2 u_L^{h}, \qquad F_{LL}^{\cos\phi_h} = - \frac{2|\vec  P_{h\perp}|}{Q} x_B^2 l_L^{h},\label{eq:FLLcosphi} \\
& F_{UT}^{\sin\phi_S} = - \frac{2M}{Q} x_B^2 u_T, \qquad\quad F_{LT}^{\cos\phi_S} = - \frac{2M}{Q} x_B^2 l_T, \label{eq:FLTcosphiS}\\
& F_{UT}^{\sin(2\phi_h-\phi_S)} = - \frac{\vec P_{h\perp}^2}{QM} x_B^2 u_{T}^{h}, \quad F_{LT}^{\cos(2\phi_h-\phi_S)} = - \frac{\vec P_{h\perp}^2}{QM} x_B^2 l_{T}^{h} \label{eq:Ftwist3-8}.
\end{align}
The remaining six structure functions are all zero up to twist-3.
We see that half of the eight twist-3 structure functions are related to the transverse polarization-dependent fracture functions.

\subsection{Azimuthal or spin asymmetries}
In addition to structure functions, one can also construct various azimuthal or spin asymmetries by
\begin{align}
\langle {\cal F} \rangle_{{\cal P}_e {\cal P}_N} \equiv \int \frac{d\sigma}{d x d y d \xi_h  d\psi d^2 P_{h\perp}} {\cal F} d\phi_h d\psi \bigg / \int \frac{d\sigma}{d x d y d \xi_h d\psi d^2 P_{h\perp}} d\phi_h d\psi ,
\end{align}
where the subscripts ${\cal P}_e = U$ or $L$ and ${\cal P}_N=U$, $L$ or $T$ denote the polarization states of the electron and the nucleon target.
From our results of the structure functions, we see clearly that there are two spin-dependent azimuthal asymmetries at twist-2.
They both depend on the nucleon transverse polarization and are given by
\begin{align}
& \langle \sin(\phi_h-\phi_S) \rangle_{UT} = \frac{|\vec  P_{h\perp}|}{2M} \frac{u_{1T}^{h}(x_B,\xi_h,P_{h\perp}) }{u_1(x_B,\xi_h,P_{h\perp}) }, \\
& \langle \cos(\phi_h-\phi_S) \rangle_{LT} = \frac{|\vec  P_{h\perp}|C(y)}{2MA(y)} \frac{l_{1T}^{h}(x_B,\xi_h,P_{h\perp}) }{u_1(x_B,\xi_h,P_{h\perp}) }.
\end{align}
Here and in the below, a summation over quark flavors, i.e., $\sum_q e_q^2\cdots$, is implicit both in the numerators and the denominators.
We note that the asymmetry $\langle \sin(\phi_h-\phi_S) \rangle_{UT}$ is of Sivers-type~\cite{Sivers:1989cc} and it does not depend on $y$ because of the cancellation of the common $A(y)$ factors associated with $F_{UT,T}^{\sin(\phi_h-\phi_S)}$ and $F_{UU,T}$ in the cross section.

We have in particular eight azimuthal or spin asymmetries at twist-3 associated with the eight twist-3 structure functions in Eqs.~(\ref{eq:Ftwist3-1})-(\ref{eq:Ftwist3-8}), i.e.,
\begin{align}
& \langle \cos\phi_h \rangle_{UU} = -\frac{|\vec  P_{h\perp}|}{Q} \frac{B(y)}{A(y)}\frac{x_B u^{h}(x_B,\xi_h,P_{h\perp}) }{u_1(x_B,\xi_h,P_{h\perp}) }, \\
& \langle \sin\phi_h \rangle_{LU} = \frac{|\vec  P_{h\perp}|}{Q} \frac{D(y)}{A(y)}\frac{x_B l^{h}(x_B,\xi_h,P_{h\perp}) }{u_1(x_B,\xi_h,P_{h\perp}) }, 
\label{eq:beamspin}\\
& \langle \sin\phi_h \rangle_{UL} = -\frac{|\vec  P_{h\perp}|}{Q} \frac{B(y)}{A(y)}\frac{x_B u_L^{h}(x_B,\xi_h,P_{h\perp}) }{u_1(x_B,\xi_h,P_{h\perp}) }, \\
& \langle \cos\phi_h \rangle_{LL} = -\frac{|\vec  P_{h\perp}|}{Q} \frac{D(y)}{A(y)}\frac{x_B l_L^{h}(x_B,\xi_h,P_{h\perp}) }{u_1(x_B,\xi_h,P_{h\perp}) }, \\
& \langle \sin\phi_S \rangle_{UT} = -\frac{M}{Q} \frac{B(y)}{A(y)}\frac{x_B u_T(x_B,\xi_h,P_{h\perp}) }{u_1(x_B,\xi_h,P_{h\perp}) }, \\
& \langle \cos\phi_S \rangle_{LT} = -\frac{M}{Q} \frac{D(y)}{A(y)}\frac{x_B l_T(x_B,\xi_h,P_{h\perp}) }{u_1(x_B,\xi_h,P_{h\perp}) }, \\
& \langle \sin(2\phi_h-\phi_S) \rangle_{UT} = -\frac{\vec P_{h\perp}^2}{2MQ} \frac{B(y)}{A(y)}\frac{x_B u_T^{h}(x_B,\xi_h,P_{h\perp}) }{u_1(x_B,\xi_h,P_{h\perp}) }, \\
& \langle \cos(2\phi_h-\phi_S) \rangle_{LT} = -\frac{\vec P_{h\perp}^2}{2MQ} \frac{D(y)}{A(y)}\frac{x_B l_T^{h}(x_B,\xi_h,P_{h\perp}) }{u_1(x_B,\xi_h,P_{h\perp}) }.
\end{align}

One can see that at the order we are considering, each azimuthal or spin asymmetry in the TFR is only generated by a specific fracture function.
This suggests that interpretations of these functions from experimental data may be simpler and more straightforward compared to the CFR at small $P_{h\perp}$, where multiple TMD PDFs and FFs are typically involved and intertwined in the asymmetry~\cite{Bacchetta:2006tn}. 
Some of the twist-3 asymmetries, such as $\langle \sin\phi_h \rangle_{UL}$ and $\langle \sin\phi_h \rangle_{LU}$, have already been measured in the TFR by CLAS12 at JLab~\cite{Harut}. 
Of particular interest is the beam-spin asymmetry $\langle \sin\phi_h \rangle_{LU}$ in Eq.~(\ref{eq:beamspin}), which is related to a twist-3 longitudinal quark fracture function $l^h$. 
A preliminary analysis shows that it undergoes a clear sign flip from the TFR to the CFR and could serve as an efficient tool to understand the transition between the production mechanisms~(Sec. 5.3 in~\cite{Accardi:2023chb}). 
Further experimental measurements will provide us with more information about the relevant fracture functions, especially the twist-3 ones.

\section{Summary}
\label{sec:Summary} 
In summary, we have derived the hadronic tensor up to twist-3 level for SIDIS with hadron production in the target fragmentation region.
The hadronic tensor at the considered order is shown to be expressed by gauge-invariant fracture functions defined with two-parton correlations.
Based on the obtained hadronic tensor, the results for structure functions are derived for both polarized lepton beam and polarized nucleon target. At the tree level, there are four structure functions at twist-2 and eight structure functions at twist-3.
Azimuthal or spin asymmetries are given based on the results of the structure functions.
These observables are all expressed using twist-2 or twist-3 collinear fracture functions.
Possible connections to experimental measurements are discussed.
Future SIDIS experiments measuring these azimuthal or spin asymmetries will provide  opportunities to extract the corresponding fracture functions.

\section*{Acknowledgments}
We would like to thank Harut Avakian for bringing our attention to this project and insightful discussions. 
We also thank Timothy Hayward for useful discussions. 
The work is supported by National Natural Science Foundation of People’s Republic of China Grants No. 12075299, No. 11821505, No. 11847612, and No. 11935017  and by the Strategic Priority Research Program of Chinese Academy of Sciences, Grant No. XDB34000000. 
K. B. Chen is supported by National Natural Science Foundation of China (Nos. 12005122, 11947055) and Shandong Province Natural Science Foundation (No. ZR2020QA082).
X. B. Tong is supported by the CUHK-Shenzhen university development fund~(No. UDF01001859) and the China Postdoctoral Science Foundation~(No. 2022M723065).

\end{document}